\documentstyle[11pt,appbwb]{article}

\input psfig
\preprint{}

\begin{document}

\title{Two Hagedorn temperatures:\\
a larger one for mesons, a lower one for baryons}
\author{Wojciech BRONIOWSKI 
\address{H. Niewodnicza\'{n}ski Institute of Nuclear Physics,
         PL-31342 Krak\'{o}w, POLAND}}
\date{}
\maketitle

\begin{abstract}
Our experimental work involved reading throught Particle Data Tables\cite
{PDG}. We show that the Hagedorn temperature is larger for mesons than for
baryons.{\em Talk presented at Meson 2000, 19-23 May 2000, Cracow, Poland}
\end{abstract}


%
%

{\em This research is being done in collaboration with Wojciech Florkowski
and Piotr \.{Z}enczykowski.} The famous Hagedorn hypothesis \cite
{hagedorn,hag94} states that at asymptotically large masses, $m$, the
density of hadronic resonance states, $\rho (m)$, behaves as 
\begin{figure}[tbp]
\centerline{%
\psfig{figure=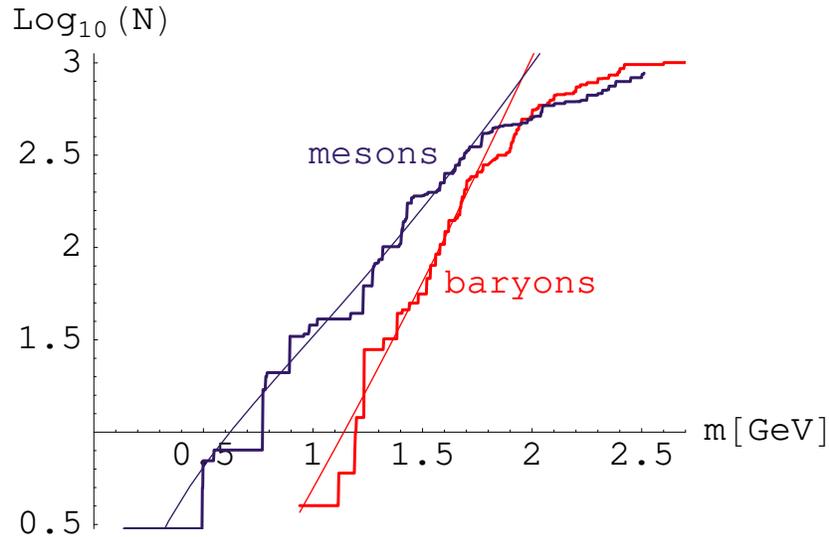,height=7.2cm,bbllx=82bp,bblly=384bp,bburx=531bp,bbury=680bp,clip=}}
\vspace{0mm}
\caption{Cumulants of meson and baryon spectra, and the Hagedorn-like fit.}
\end{figure}
\begin{equation}
\rho (m)\sim \exp \left( \frac{m}{T_{H}}\right)  \label{hag}
\end{equation}
The Hagedorn temperature, $T_{H}$, is a scale controlling the exponential
growth of the spectrum.\footnote{$T_{H}$ need not immediately be associated
with thermodynamics, here we are just concerned with the spectrum of
particles {\em per se}.} Ever since hypothesis (\ref{hag}) was made, it has been believed
that there is one universal $T_{H}$ for all hadrons. {\em Presently
available experimental data show that this is not the case} \cite{twoT}.

In Fig. 1 we compare the {\em cumulants} of the spectrum, defined as the
number of states with mass lower than $m$. The experimental curve is $N_{%
{\rm exp}}(m)=\sum_{i}g_{i}\Theta (m-m_{i})$, where $%
g_{i}=(2J_{i}+1)(2I_{i}+1)$ is the spin-isospin degeneracy of the $i$th
state, and $m_{i}$ is its mass. The theoretical curve corresponds to $N_{%
{\rm theor}}(m)=\int_{0}^{m}\rho _{{\rm theor}}(m^{\prime })dm^{\prime }$,
where $\rho _{{\rm theor}}(m)=f(m)\exp (m/T)$, with $f(m)$ denoting a
slowly-varying function. A typical choice \cite{hag94}, used in the plot, is 
$f(m)=A/(m^{2}+(500{\rm MeV})^{2})^{5/4}$. Parameters $T_{H}$ and $A$ are
obtained with the least-square fit to $\log N_{{\rm theor}}$. Other choices
of $f(m)$ give fits of similar quality. A striking feature of Fig. 1 is the
linearity of $\log N$ starting at very low $m$, and extending till $m\sim 1.8%
{\rm GeV}$. Clearly, this shows that (\ref{hag}) is valid in the range of
available data.\footnote{%
Above 1.8GeV the data is sparse and we have to wait for this region to be
filled in by future experiments.} However, the slopes in Fig. 1 are
different for mesons and baryons. For the assumed $f(m)$ we get $T_{{\rm %
meson}}=195$MeV and $T_{{\rm baryon}}=141$MeV. This means that $T_{{\rm meson%
}}>T_{{\rm baryon}}$, and the inequality is substantial. Although it has
been known to researchers in the field of hadron spectroscopy that the
baryons multiply more rapidly than mesons, to our knowledge this fact has
not been presented as vividly as in Fig. 1. To emphasize the strength of the
effect we note that in order to make the meson line parallel to the baryon
line, we would have to aggregate $\sim 500$ additional meson states up to $%
m=1.8$MeV as compared to the present number of $\sim 400$. If Ref. \cite
{twoT} we show that the fitted values of $T_{H}$, distinct for mesons and
baryons, do not depend on flavor.

Why do mesons and baryons behave so differently? First, let us stress that
it is not easy to get an exponentially rising spectrum at all. Take the
simplistic harmonic-oscillator model, whose density of states grows as $%
m^{d-1}$, with $d$ denoting the number of dimensions. For mesons there is
one relative coordinate, hence $\rho \sim m^{2}$, whereas the two relative
coordinates in the baryon give $\rho \sim $ $m^{5}$. Weaker-growing
potentials lead to a faster growth, but fall short of the behavior (\ref{hag}%
). We know of two approaches yielding behavior (\ref{hag}), both involving
combinatorics of infinitely-many degrees of freedom. {\em Statistical
bootstrap} models \cite{hagedorn,SBM2} form particles form clusters of
particles, and employ the principle of self-similarity. It can be shown,
following {\em e.g.} the steps of Ref. \cite{Nahm}, that the model leads to
equal Hagedorn temperatures for mesons and for baryons.\footnote{%
Since baryons are formed by attaching mesons to the ``input'' baryon, the
baryon spectrum grows at exactly the same rate as the meson spectrum.} Thus
the bootstrap idea {\em is not capable} of explaining the different behavior
of mesons and baryons in Fig. 1.

On the other hand, the {\em Dual String models }\cite{Jacob} do give the
demanded effect of $T_{{\rm meson}}>T_{{\rm baryon}}$, at least at
asymptotic masses. Let us analyze mesons first. The particle spectrum is
generated by the harmonic-oscillator operator describing vibrations of the
string, $N$ $=\sum_{k=1}^{\infty }\sum_{\mu =1}^{D}ka_{k,\mu }^{\dagger
}a_{k,\mu }$, where $k$ labels the modes and $\mu $ labels additional
degeneracy, related to the number of dimensions \cite{Jacob}. Eigenvalues of 
$N$ are composed in order to get the square of mass of the meson, according
to the formula $\alpha ^{\prime }m^{2}-\alpha _{0}=n,$ where $\alpha
^{\prime }\sim 1{\rm GeV}^{-2}$ is the Regge slope, and $\alpha _{0}\approx 0
$ is the intercept. Example: take $n=5$. This can be made by taking the $k=5$
eigenvalue of $N$ (this is the leading Regge trajectory, with maximum
angular momentum), but we can also take obtain the same $m^{2}$ by exciting
one $k=4$ and one $k=1$ mode, alternatively $k=3$ and $k=2$ modes, {\em etc.}
The number of possibilities corresponds to partitioning the number $5$ into
natural components: 5, 4+1, 3+2, 3+1+1, 2+2+1, 2+1+1+1, 1+1+1+1+1.
Partitions with more than one component describe the sub-leading Regge
trajectories. With $D$ degrees of freedom each component can come in $D$
different species. Let us denote the number of partitions in our problem as $%
P_{D}(n)$. For large $n$ the asymptotic formula for {\em partitio numerorum}
leads to the exponential spectrum according to the formula \cite
{partitio,Jacob}. 
\begin{equation}
\rho(m)=2\alpha ^{\prime }mP_{D}(n),\quad P_{D}(n)\simeq \sqrt{\frac{1}{%
2n}}\left( \frac{D}{24n}\right) ^{\frac{D+1}{4}}\exp \left( 2\pi \sqrt{\frac{%
Dn}{6}}\right) ,  \label{parti}
\end{equation}
where $n=\alpha ^{\prime }m^{2}$. We can now read-off the mesonic Hagedorn
temperature: $T_{{\rm meson}}=\frac{1}{2\pi }\sqrt{\frac{6}{D\alpha ^{\prime
}}}$. 
\begin{figure}[tbp]
\centerline{%
\psfig{figure=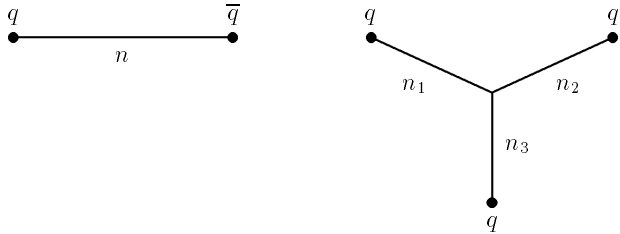,height=4.0cm,bbllx=149bp,bblly=529bp,bburx=457bp,bbury=665bp,clip=}}
\vspace{-2mm}
\caption{Meson and baryon string configurations.}
\end{figure}
Now the baryons: the configuration for the baryon is shown in Fig. 2. The
three strings vibrate independently, and the corresponding vibration
operators, $N$, add up. Consequently, their eigenvalues $n_{1}$, $n_{2}$,
and $n_{3}$ add up. Thus we simply have a partition problem with $3$ times
more degrees of freedom than in the meson. The replacement $D\rightarrow 3D$
in (\ref{parti}) leads immediately to $T_{{\rm baryon}}=\frac{1}{2\pi }\sqrt{%
\frac{2}{D\alpha ^{\prime }}}$, $T_{{\rm meson}}/T_{{\rm baryon}}=\sqrt{3}$.
We stress the picture is fully consistent with the Regge phenomenology. The
leading Regge trajectory for baryons is generated by the excitation of a
single string, {\em i.e. } two out of three numbers $n_{i}$ vanish. This is
the quark-diquark configuration. The subleading trajectories for baryons
come in a much larger degeneracy than for mesons, due to more combinatorial
possibilities. The slopes of the meson and baryon trajectories are
universal, and given by $\alpha ^{\prime }$. We stress that the
``number-of-strings'' mechanism described above is asymptotic. When applied
to the data in the observed region one can, however, obtain very good
agreement for mesons with a wide range of the dimensionality parameter $D$ 
\cite{dienes}. Baryon slopes can also be descibed properly, however with
simplest string models there are too many baryon states. This hints for
improvements, {\em e.g.} the inclusion of the spin-flavor symmetry factors
for baryon states. More work needs to be done here.

We summarize the basics of the string mechanism: as $m^{2}$ increases, more
and more degrees of freedom ``wake up''. Via partitio numerorum they lead to
the exponential growth of the spectrum. The three strings in the baryon
bring more degrees of freedom and result in ``faster'' combinatorics. We
have heard many talks in this conference on hadron exotics. If an exotic is
a multi-string configuration (generalizations of Fig. 2), then the
corresponding spectrum will grow exponentially with the Hagedorn temperature
inversely proportional to the square root of the number of strings. For
instance, $T_{q\overline{q}q\overline{q}}=\frac{1}{2}T_{{\rm meson}}$. This
is reminiscent of the effect described in Ref. \cite{cudell}.

The author thanks Keith R. Dienes for many profitable e-mail discussions on
the issues of hadron spectra and string models, and to Andrzej 
Bia\l{}as and Kacper Zalewski for numerous useful comments.


\end{document}